# The role of quantum interference in determining transport properties of organic molecules

Kamil Walczak [1]

Institute of Physics, Adam Mickiewicz University
Umultowska 85, 61-614 Poznań, Poland

An analytic approach to the electron transport phenomena in molecular devices is presented. Analyzed devices are composed of organic molecules attached to the two semi-infinite electrodes. Molecular system is described within the tight-binding model, while the coupling to the electrodes is analyzed through the use of Newns-Anderson chemisorption theory. The current-voltage (I-V) characteristics are calculated through the integration of transmission function in the standard Landauer formulation. The essential question of quantum interference effect of electron waves is discussed in three aspects: (i) the geometry of molecular bridge and (ii) the presence of an external magnetic field and (iii) the strength of the molecule-to-electrodes coupling.

Key words: molecular device, quantum interference, electronic transport, antiresonance, molecular electronics

PACS numbers: 85.65.+h , 73.23.-b

## I. Introduction

Recent advances in fabrication of molecular junctions based on single molecules [1-3] or molecular strands [4-6] have attracted much attention in the experimental and theoretical study of these nanoscopic systems. As a matter of fact molecular (or atomic) level represents the ultimate in device miniaturization. The operation of such two-terminal devices is due to an applied bias. The current flowing across the junction is strongly nonlinear function of bias voltage and its detailed description is a very complex problem. The full knowledge of the conduction mechanism on this scale is not well understood yet, but the transport properties of these systems are associated with some effects of quantum nature (such as: quantization of molecular energy levels, quantum interference of electron waves and discreteness of electron charge and spin). A quantitative understanding of the physical mechanisms underlying the operation of molecular-scale devices remains a major challenge in nanoelectronics research.

The main aim of this paper is to introduce an analytic approach, based on tight-binding model, to study electronic transport in molecular devices. Although some *ab initio* methods were used to calculate the conductance [7-12], there is still the need of doing it in a simple parametric approaches [13-17], especially in the case of a bigger molecular systems. The parametric study is motivated by the fact that *ab initio* theories are computationally quite expensive and here we focus our attention on qualitative effects rather (some tendencies) then the quantitative ones. That's why we restrict our considerations only to a simple analytical formulation of the transport problem.

As a starting point we consider model of benzene-1,4-dithiolate (BDT) molecule [2] in between two semi-infinite electrodes. Theoretical investigations presented here taking into account the possibility of various geometries of molecular bridge and the presence of an external magnetic field.

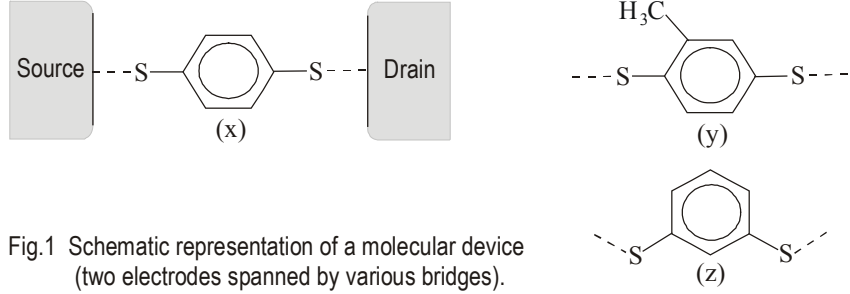

Fig.1 Schematic representation of a molecular device
(two electrodes spanned by various bridges).

## II. A glimpse onto the theoretical formulation

Molecular device is considered as a system of organic molecule sandwiched between two metallic electrodes (see Fig.1). Since π-electron subsystem dictates the transport properties of organic molecules, we have decided to describe molecular bridge through the use of the single-band tight-binding model defined with the help of one-electron Hamiltonian:

$$H = \sum_{i,\sigma} \varepsilon_{i,\sigma} c^+_{i,\sigma} c_{i,\sigma} - \sum_{i,\sigma} \left( c^+_{i,\sigma} t_{i,i+1} c_{i,\sigma} + h.c. \right), \qquad (1)$$

where: $\varepsilon_{i,\sigma} \equiv \varepsilon$ is the local site energy, $t_{i,i+1} \equiv t$ is the first-neighbour hopping integral, and $c^+_{i,\sigma}$, $c_{i,\sigma}$ are creation and annihilation operators for an electron on site i with spin $\sigma$, respectively. In our simplified picture, the current is depicted as a single-electron scattering process between two reservoirs of charge carriers. The molecule itself acts as a strong defect in the system and the current flowing through the device is computed as the total transfer rate (multiplied by an elementary electronic charge) [18]:

$$I(V) = \frac{2e}{h} \int_{-\infty}^{+\infty} T(E,V) [f(E - \mu_S) - f(E - \mu_D)] dE, \qquad (2)$$

where: $f(x)$ is the standard Fermi distribution function (in the zero temperature limit it becomes the step function), $\mu_S$ and $\mu_D$ are the electrochemical potentials of the source and drain, respectively. In our non-self-consistent approach we must postulate potential distribution along the molecule. For the sake of simplicity we assume that the electric field inside the molecule have a minimal effect on the current-voltage spectrum, so the voltage is dropped entirely at the molecule/electrode interfaces [19]:

$$\mu_{S/D} = E_F \pm eV/2. \qquad (3)$$

($E_F$ is Fermi energy level of the electrodes). Anyway, the shape of potential profile do not change our general conclusions.

The transmission function represents the ease with which electrons can be transferred from the source electrode, through the molecular bridge, to the drain and it is expressed as follows [18]:

$$T(E,V) = tr\left[ (Q_S - Q^+_S) G (Q^+_D - Q_D) G^+ \right]. \qquad (4)$$

The Green function of the molecule is given through the relation:

$$G = [ES - H - Q_S - Q_D]^{-1}, \qquad (5)$$

were S is the overlap matrix (which is equal to identity matrix for the orthogonal basis set of states). All the information concerning the molecule-to-electrodes coupling is included into the self-energies $Q_S$ and $Q_D$. These functions describe the molecular eigenstates broadening (by imaginary parts) and the energy shift (by real parts) that the molecule experiences upon the electrode binding. Mentioned coupling is treated within the Newns-Anderson chemisorption model [13,20]:

$$Q_{S/D} = \frac{\beta^2}{\gamma}\left[\frac{E \pm eV/2}{2\gamma} - i\sqrt{1 - \left(\frac{E \pm eV/2}{2\gamma}\right)^2}\right], \quad (6)$$

where: E is the injection energy of the tunneling electron, $\beta$ is the hopping parameter between the surface of each of the electrodes to molecular bridge and $4\gamma$ is the electron reservoir energy bandwidth. In this model, the voltage dependence of the transmission function is negligible and for analyzed range of applied voltages it has very little effect on the conduction.

### III. Results and their interpretation

In this paper we concentrate ourselves on transport characteristics (transmission-energy and current-voltage functions) of a variety of the molecular devices. As a starting point we consider the sample consisted of BDT molecule attached to ideal electrodes (as shown in Fig.1 with x-bridge) at the room temperature (293 K). This is a test case simple enough to modify and analyze all the essential features of the problem in detail. Common set of energy parameters (given in eV) used in our calculations (throughout this work) are chosen as follows: $\varepsilon = 0$ (the reference energy), $t = 2.5$, $\beta = 0.5$ (weak-coupling case) or $\beta = 2.0$ (strong-coupling case), $\gamma = 10.0$ [20]. Another essential problem is associated with the precise location of the Fermi level relatively to the molecular energy levels. For simplicity, we make a realistic assumption that Fermi energy is placed in the middle of the HOMO-LUMO gap ($E_F = 0$) [7].

Here we briefly summarize all the essential features of the transport characteristics in two distinguished regimes: weak ($\beta \ll t$) and strong ($\beta \sim t$) molecule-to-electrodes couplings. In the weak-coupling case, the total transmission is almost everywhere very small, except at the resonances where it approaches unity (see Fig.2a). Such peaks in the transmission spectrum coincide with eigenenergies of the isolated molecule, so the transmission itself reproduces the electronic structure of the molecular bridge. With the increase of the strength of the coupling, the resonances are shifted and gain substantial widths (see Fig.2b).

The scenario for electron transfer through the molecular junction seems to be very simple. As the voltage is increased, the electrochemical potentials on the electrodes are shifted and eventually cross one of the molecular energy level. It opens a current channel, what is observed as a jump in the I-V dependence. The shape and height of the current steps depend on the width of molecular resonances. In particular, the height of a step in the I-V curve is directly proportional to the area of the corresponding peak in the transmission spectrum (since the current is obtained from the transmission function in the integration procedure). In the weak-coupling case, the I-V characteristic shows a staircase-like structure with sharp steps [20], which is associated with discrete nature of molecular resonances (see Fig.3a). With the increase of the strength of the coupling, the current turns into continuous function of bias voltage and achieves much more bigger values (see Fig.3b).

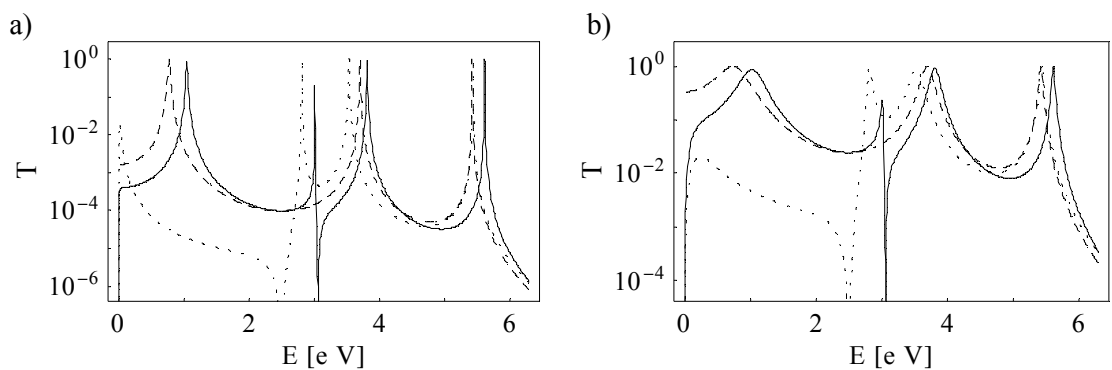

Fig.2 Transmission $T(E) = T(-E)$ as a function of electron energy (with respect to Fermi energy level) for devices with bridges: x (broken line), y (solid line) and z (dotted line) in the case of weak (a) and strong (b) coupling with the electrodes, respectively.

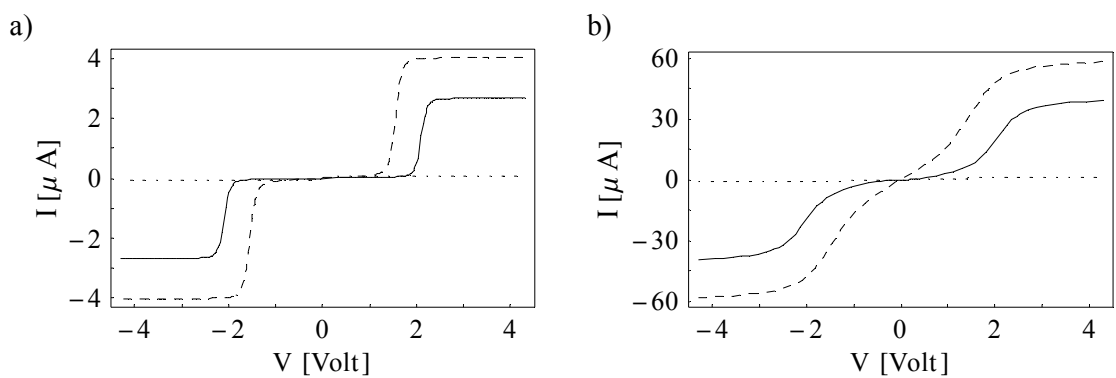

Fig.3 Comparison of the current-voltage characteristics for devices with bridges: x (broken line), y (solid line) and z (dotted line) in the case of weak (a) and strong (b) coupling with the electrodes, respectively.

**A. The geometry of a molecular bridge**

As an example we consider three different bridges (x, y and z depicted schematically in Fig.1) to take into account some differences in the geometry of the molecule relatively to BDT. The y-bridge is referred to the physical situation in which we add a chemical substituent group ($-CH_3$), which can be treated as typical π-electron compound (Of course, different substituents could be used and modeled as heteroatoms [21]). The z-bridge is associated with the variation of the character of molecular binding to the electrodes (with respect to the initial BDT symmetric case). The most important thing is that in this way the interference conditions are changed.

In considered nanoscopic devices, the electrons are carried from the source to the drain through benzene-like molecules. So electron waves propagating along two branches of the ring may suffer a phase shift between themselves (due to result of quantum interference between the various pathways that the electron can take [22,23]). Such effects lead to strengthening or weakening the amplitude of the probability of the electron appearance behind the ring (according to the standard interpretation of the wave function). It manifests itself especially as transmittance cancellations (some peaks do not reach unity anymore) and antiresonances in the transmission spectrum.

Here mentioned phase shift is realized by the variation of the geometry of a molecular bridge. For symmetrical BDT molecule we do not observe two states ($E = \pm 2.500$ eV), which are uncoupled with the states associated with the transmission function (broken line in Fig.2). Breaking of the molecular symmetry results not only in shifting the perpendicular resonances, but also in coupling ("mixing") with unrevealed states, and therefore more peaks are present in the transmission spectrum (solid and dotted lines in Fig.2). Such additional resonances do not contribute to the I-V characteristics, because their widths are very narrow and have negligibly small effect on the integration procedure (see Fig.3).

Another remarkable feature of the transmission is the existence of the transmittance zeros [22-24]. Such antiresonance states are specific to the interferometric nature of the scattering and do not occur in conventional one-dimensional scattering problems of potential barriers [25]. In the case of y-bridge there are three antiresonances for electron energies: $E = 0$ eV and $E = \pm 3.062$ eV, while for z-bridge we have obtained: $E = 0$ eV and $E = \pm 2.500$ eV (see Fig.2). Positions of antiresonances on the energy scale are independent from the strength of the molecule-to-electrodes coupling. The widths of that states are again very narrow and can not be observed in the I-V dependences. However, the variations of interference conditions have influence on the magnitude of the current flowing through the system (see Fig.3).

**B. The presence of an external magnetic field**

Now we proceed to control the phase shift by applying an external magnetic field. As an example, we consider initial configuration of the BDT-based device (x-bridge in Fig.1), where benzene ring is threaded by magnetic field perpendicular to the plane of the molecule. Magnetic flux penetrating the ring is incorporated in the hopping parameters through the use of the Peierls gauge phase factors [26-30]:

$$t_{i,i+1} = t \exp\left[-i\frac{\pi}{N}\frac{\phi}{\phi_0}\right] \qquad (7)$$

where: $2N$ is the number of atoms in the ring, $\phi = \int \vec{B} \cdot d\vec{S}$ is magnetic flux crossing the ring and $\phi_0 = h/e$ ($\sim 4.136 \times 10^{-15}$ Wb) is an elementary quantum of flux. From the above expression (7), one can see that the electron gains an additional phase each time it hops from one site of the ring to another (sign of that phase depends on the hopping direction with respect to the field). According to the phase shift, there can be a constructive or destructive interference after the electron propagation through the benzene ring.

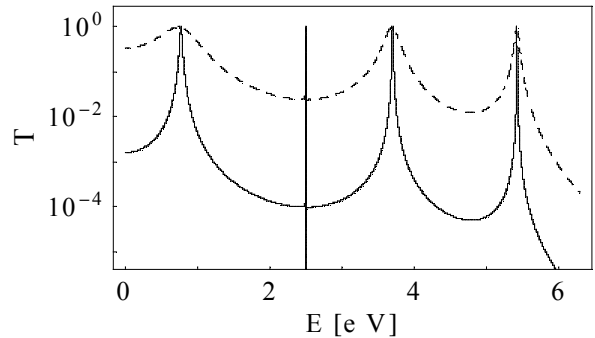

Fig.4 Transmission $T(E) = T(-E)$ as a function of electron energy (with respect to Fermi energy level) for BDT-based device in the presence of magnetic field $B = 40$ T in the case of weak (solid line) and strong (broken line) coupling with the electrodes, respectively.

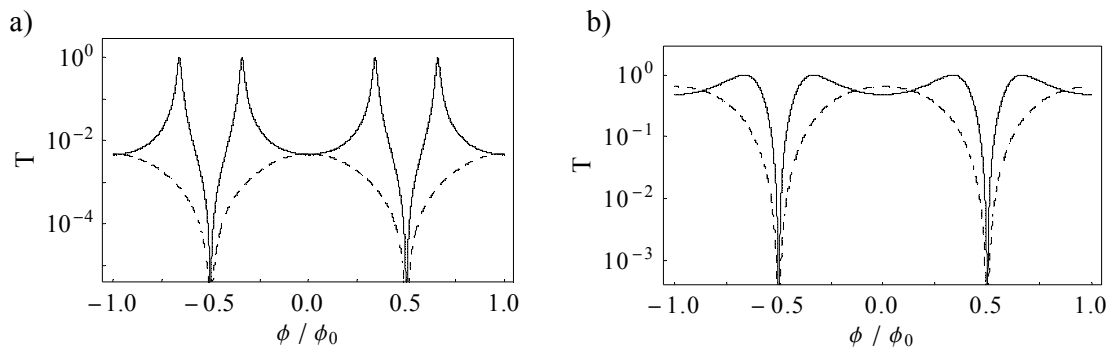

Fig.5 Transmission as a function of magnetic flux for BDT-biased device and the injection energy of the tunneling electron $E = 0.5$ eV (solid line) and $E = 1.0$ eV (broken line) in the case of weak (a) and strong (b) coupling with the electrodes, respectively.

In Fig.4 we plot transmission as a function of electron energy for BDT-based device, where magnetic field $B = 40$ T penetrates the benzene ring. Here we can also observe the activation of the "hidden" states of the molecular bridge as two additional peaks (at energies $E = \pm 2.4995$ eV). Each of resonances reach unity and are accompanied by antiresonances (at energies $E = \pm 2.500$ eV). Furthermore, Fig.5 shows the Aharonov-Bohm oscillations in the transmission as a function of the magnetic flux $\phi$ encircled by the ring. As expected, the transmission spectrum is flux periodic for all energies of tunneling electron with a period of $\phi_0$. However, there are some suggestions that in the presence of impurities the period is changed to $\phi_0 / 2$ [31,32]. In Fig.5 we can also see that transmission decreases to zero for the two values of magnetic flux: $\phi = \pm \phi_0 / 2$. Such states are again coupling-independent and $\phi_0$-repeated in a periodical way.

Although there are some modifications of the transmission function due to the application of a transverse magnetic field, there are no changes in the I-V characteristics. Therefore, the current of two-terminal molecular device can not be controlled by a transverse magnetic field, which seems to be too large to produce observable phase shift (according to our simulations – hundreds or even thousands of Teslas).

**C. The influence of a chemical substituent**

The purpose of this subsection is to make some predictions regarding to the influence of the substituent's location on transport properties of benzene chains. As an example, we take into consideration a linear chain of three benzene rings coupled to the electrodes with the help of thiol groups (see Fig.6). The chemical substituent ($-CH_3$) can be attached to one of this ring in the positions: (1) or (2), respectively. Since in this work we are focused on the interference effects only, our further analysis do not touch the problem of charge redistribution due to electronegativity of that substituent group (its releasing or withdrawing character).

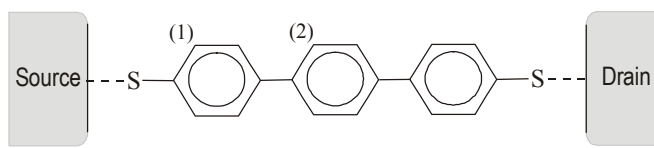

Fig.6 Schematic representation of an analyzed sample (numbers denote the position of substituent group).

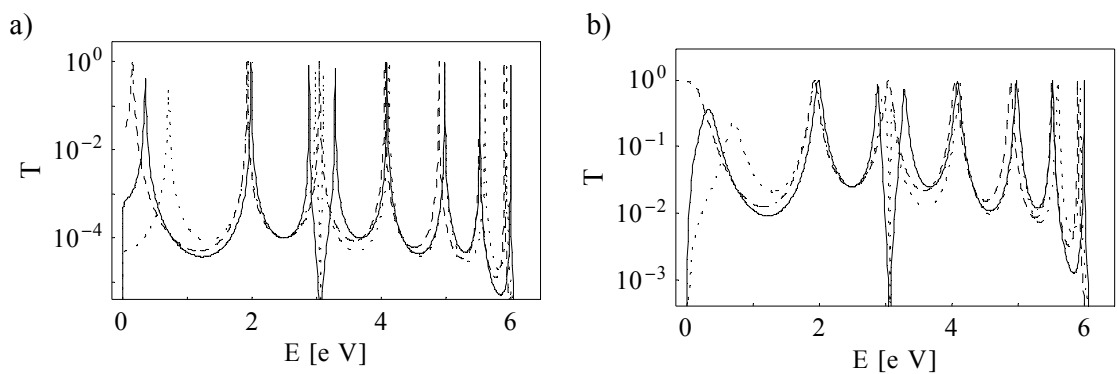

Fig.7 Comparison of the transmission functions $T(E) = T(-E)$ for three benzene rings: without substituent – broken line, and with substituent in the position (1) – dotted line, and (2) – solid line, in the case of weak (a) and strong (b) coupling with the electrodes, respectively.

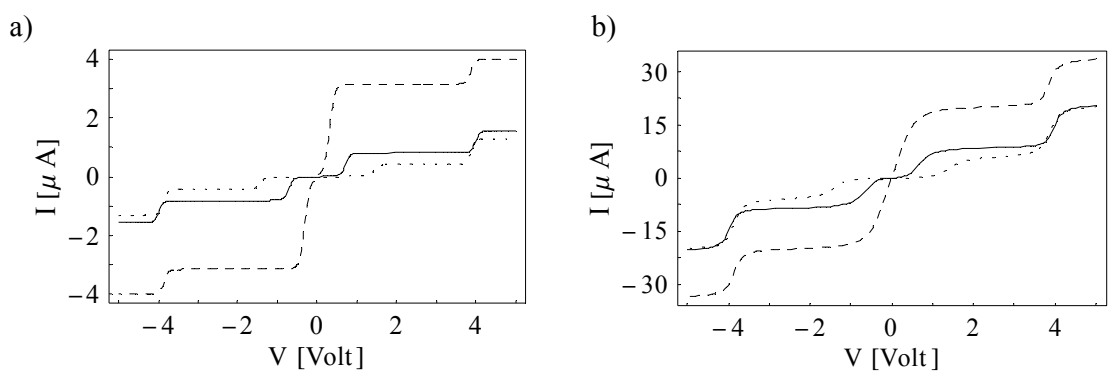

Fig.8 Comparison of the current-voltage characteristics for three benzene rings: without substituent – broken line, and with substituent in the position (1) – dotted line, and (2) – solid line, in the case of weak (a) and strong (b) coupling with the electrodes, respectively.

Transmission spectra (with respect to electron energy) are presented in Fig.7. Particularly, the most significant changes for the analyzed samples can be seen in the vicinity of the Fermi level ($E_F = 0$ eV). In the non-substituent case, central peaks reach the unity, while in the presence of the methane groups, these central peaks are shifted, lowered in height (below unity) and finally separated by antiresonance state (exactly at energy $E = 0$ eV). Since the transmission next to the Fermi energy level has the greatest impact on the I-V characteristics of molecular devices, we expect to observe the influence of the substituent's presence on the I-V dependences. Indeed, the magnitude of the current flowing through the device consisted of a non-substituent molecule is at least twice (in the strong-coupling limit) or even more (in the weak-coupling limit) as large as in the case of attached group.

Interesting behavior of the transmission functions is found in the regions around $E \sim \pm 3$ eV. One resonant state in the non-substituent system is split into two different peaks separated by antiresonances (at energies $E = \pm 3.062$ eV) in the presence of substituent group. Generally, the location of the chemical substituent group and the strength of the molecule-to-electrodes coupling have no influence on the appearance of antiresonances in the considered devices (their energies are always the same). The last conclusion is a simple confirmation of our assumption, that we have to do with quantum interference effects. However, transmission zeros are again too narrow to be distinguished in the I-V curves (being entirely blurred).

## IV. Final remarks

In this paper we introduced parametric approach, based on tight-binding model, to study the transport properties of the benzene-ring family of molecules, which can be located into an external magnetic field. In fact, presented scheme can be applied to much more complicated organic structures, which contain a special class of substituents (π-electron compounds). The influence of all two considered factors (the geometry of molecular bridge and the penetration by the magnetic field) on electronic transport phenomena seems to have the same source. It is explained as the effect of quantum interference of electron waves. Especially the origin of antiresonances in the transmission function is understood as a strict confirmation of the mentioned effect.

For the sake of simplicity, described method is based on several unrealistic assumptions. Further investigations are expected to take into account the charge transfer across the metal/molecule interfaces just upon binding to the electrodes. It creates a potential barrier for moving electrons (Schottky-like) that can modify the transport mechanism through the junction [33]. Here we also neglected possible changes in the electronic structure of the molecular bridge upon the influence of an applied bias (the static Stark effect). Such field-induced modification of the molecular states becomes important as the bias voltage increases beyond the linear-transport regime [33,34]. However, these effects can be included into our framework by a simple generalization of the presented formalism.

In the frames of this work we also ignored the influence of inelastic processes and electron correlations on the transport phenomena in molecular devices.

## Acknowledgments

The author is grateful to T. Kostyrko, B. Bułka and W. Babiaczyk for numerous helpful discussions. Work supported in part by K.B.N. Poland, project number 2 PO3B 087 19.

## References

[1] E-mail address: walczak@amu.edu.pl
[1]  A. Yazdani, D. M. Eigler, N. D. Lang, Science **272**, 1921 (1996).
[2]  M. A. Reed, C. Zhou, C. J. Muller, T. P. Burgin, J. M. Tour, Science **278**, 252 (1997).
[3]  R. H. M. Smit, Y. Noat, C. Untiedt, N. D. Lang, N. C. van Hemert,
     J. M. van Ruitenbeek, Nature **419**, 906 (2002).
[4]  L. A. Bumm, J. J. Arnold, M. T. Cygan, T. D. Dunbar, T. P. Burgin, L. Jones II,
     D. L. Allara, J. M. Tour, P. S. Weiss, Science **271**, 1705 (1996).
[5]  R. P. Andres, J. O. Bielefeld, J. I. Henderson, D. B. Janes, V. R. Kolagunta,
     C. P. Kubiak, W. J. Mahoney, R. G. Osifchin, Science **273**, 1690 (1997).
[6]  J. Chen, M. A. Reed, A. M. Rawlett, J. M. Tour, Science **286**, 1550 (1999).
[7]  S. N. Yaliraki, A. E. Roitberg, C. Gonzalez, V. Mujica, M. A. Ratner,
     J. Chem. Phys. **111**, 6997 (1999).
[8]  M. Di Ventra, S. T. Pantelides, N. D. Lang, Phys. Rev. Lett. **84**, 979 (2000).
[9]  Y. Xue, S. Datta, M. A. Ratner, J. Chem. Phys. **115**, 4292 (2001).
[10] J. Taylor, H. Gou, J. Wang, Phys. Rev. B **63**, 245407 (2001).
[11] P. A. Derosa, J. M. Seminario, J. Phys. Chem. B **105**, 471 (2001).
[12] P. S. Damle, A. W. Ghosh, S. Datta, Phys. Rev. B **64**, R201403 (2001);
     Chem. Phys. **281**, 171 (2002).
[13] V. Mujica, M. Kemp, M. A. Ratner, J. Chem. Phys. **101**, 6849 (1994);
     *ibid*. **101**, 6856 (1994).

[14] M. P. Samanta, W. Tian, S. Datta, J. I. Henderson, C. P. Kubiak, Phys. Rev. B **53**, R7626 (1996).
[15] E. G. Emberly, G. Kirczenow, Phys. Rev. B **58**, 10911 (1998).
[16] M. Hjort, S. Staftröm, Phys. Rev. B **62**, 5245 (2000).
[17] T. Kostyrko, J. Phys.: Condens. Matter **14**, 4393 (2002).
[18] S. Datta, *Electronic transport in mesoscopic systems*, Cambridge University Press, Cambridge 1997.
[19] W. Tian, S. Datta, S. Hong, R. Reifenberger, J. I. Henderson, C. P. Kubiak, J. Chem. Phys. **109**, 2874 (1998).
[20] V. Mujica, M. Kemp, A. E. Roitberg, M. A. Ratner, J. Chem. Phys. **104**, 7296 (1996).
[21] A. Sadlej, *Elementary Methods of Quantum Chemistry*, PWN, Warszawa 1966 (in Polish).
[22] J. M. Lopez-Castillo, A. Filali-Mouhim, J. P. Jay-Guerin, J. Phys. Chem. **97**, 9266 (1993).
[23] A. Cheong, A. E. Roitberg, V. Mujica, M. A. Ratner, J. Photochem. Photobiol. **82**, 81 (1994).
[24] E. G. Emberly, G. Kirczenow, J. Phys.: Condens. Matter **11**, 6911 (1999).
[25] L. D. Landau, E. M. Lifszitz, *Quantum Mechanics*, PWN, Warszawa 1975 (in Polish).
[26] J. P. Carini, K. A. Multtalib, S. R. Nagel, Phys. Rev. Lett. **53**, 102 (1984).
[27] M. A. Davidovich, E. V. Anda, Phys. Rev. B **50**, 15 453 (1994).
[28] M. A. Davidovich, E. V. Anda, J. R. Iglesias, G. Chiappe, Phys. Rev. B **55**, R7335 (1997).
[29] K. Haule, J. Bonča, Phys. Rev. B **59**, 13 087 (1999).
[30] A. Y. Smirnov, N. J. M. Horing, L. G. Mourokh, Appl. Phys. Lett. **77**, 2578 (2000).
[31] B. Pannetier, J. Chausay, R. Rarnmal, P. Gandit, Phys. Rev. Lett. **53**, 718 (1984).
[32] C. O. Umbach, C. van Hasendock, R. B. Laibowitz, R. B. Washburn, R. A. Webb, Phys. Rev. Lett. **54**, 2696 (1985).
[33] N. D. Lang, Ph. Avouris, Phys. Rev. Lett. **84**, 358 (2000).
[34] V. Mujica, A. E. Roitberg, M. A. Ratner, J. Chem. Phys. **112**, 6834 (2000).